\documentclass[11pt,twoside]{article}

\usepackage{asp2006}
\usepackage{epsf}
\usepackage{psfig}
\usepackage{lscape}

\begin{document}
\title{A CCD Photometric Study and Search for Pulsations in RZ Dra and EG Cep}   
\author{R. Pazhouhesh$^{1}$, A. Liakos$^{2}$ and P. Niarchos$^{2}$}   
\affil{$^{1}$ Physics Department, Faculty of Science, University of
Birjand, Birjand, Iran\\
$^{2}$ Department of Astrophysics,Astronomy and Mechanics, Faculty
of Physics, National and Kapodistrian University of Athens,
Athens, Greece}

\begin{abstract}
This paper presents CCD observations of the Algol-type eclipsing
binaries RZ Dra and EG Cep. The light curves have been analyzed
with the PHOEBE software and Wilson-Devinney code (2003 version).
A detailed photometric analysis, based on these observations, is
presented for both binarity and pulsation. The results indicate
semidetached systems where the secondary component fills its Roche
lobe. After the subtraction of the theoretical light curve, a
frequency analysis was performed in order to check for pulsations
of the primary component of each system. Moreover, a period
analysis was performed for each case in order to search for
additional components around the eclipsing pairs.
\end{abstract}

\section{Introduction}
Eclipsing Binaries are important astrophysical objects and several
of them are known to contain a pulsating component. It is
interesting to study such systems, since extra information can be
extracted from both their pulsation and eclipsing properties,
leading to a more reliable determination of system parameters.\\
EG Cep and RZ Dra are referred as candidate systems for including
pulsating components \citep{S06} and therefore have been selected
for observations and study.

\section{Observations}
The photometric observations were carried out at the
Gerostathopoulion Observatory of the University of Athens during
12 nights from May to August 2007, using the 0.4m Cassegrain
telescope equipped with the ST-8XMEI CCD camera and BVRI Bessell
photometric filters. The comparison and check stars used for each
system are presented in Table \ref{tab1} and the calculated times
of minima obtained from our observations, using the method of
\citet{KVW56}, are given in Table \ref{tab2}.

\begin{table}
\caption{The photometric observations log}
\label{tab1}
\begin{center}
\scalebox{0.7} {
\begin{tabular}{ccccc}

\hline
\textbf{System}& \textbf{Nights spent} & \textbf{Filters used}&   \textbf{Comparison Star}  &     \textbf{Check  Star}   \\
\hline
    EG Cep     &            7          &        BVRI          &         GSC 4589:2757       &       GSC 4589:2842        \\
    RZ Dra     &            5          &         RI           &         GSC 3916:1889       &       GSC 3916:1825        \\
\hline
\end{tabular}}
\end{center}
\end{table}

\begin{table}
\caption{The times of minima derived from our observations}
\label{tab2}
\begin{center}
\scalebox{0.7} {
\begin{tabular}{ccc}

\hline
\textbf{System} & \textbf{HJD-2400000.0}& \textbf{Filters}  \\
\hline
EG Cep          &     54264.5537~(1)    &        RI         \\
                &     54265.3706~(2)    &        RI         \\
                &     54334.5376~(3)    &       BVRI        \\
RZ Dra          &     54201.5765~(1)    &         R         \\
                &     54229.3945~(1)    &        RI         \\
                &     54232.4247~(2)    &        RI         \\
\hline
\end{tabular}}
\end{center}
\end{table}

\section{O-C Analysis}
\begin{figure}[h]
\plottwo {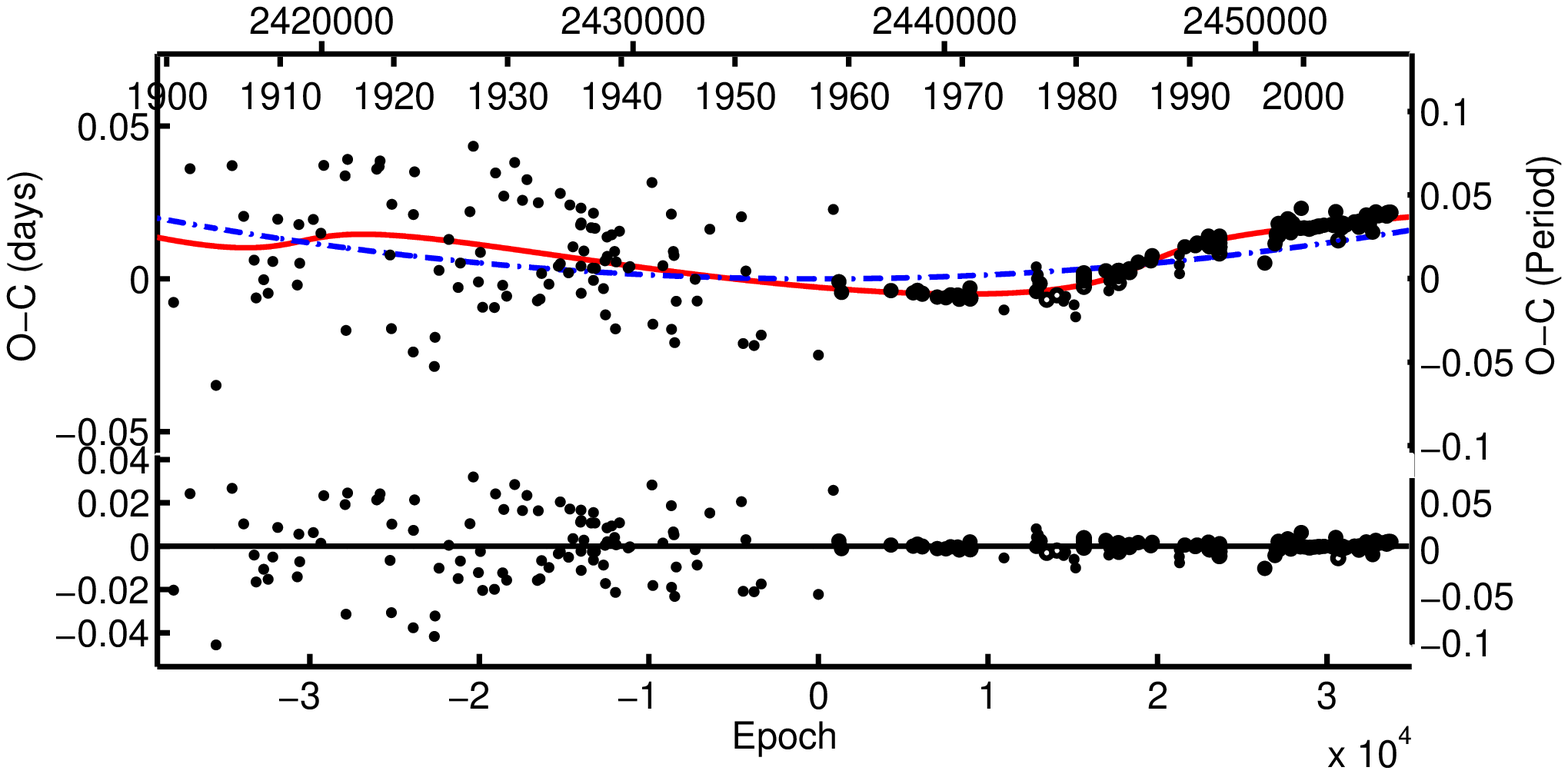}{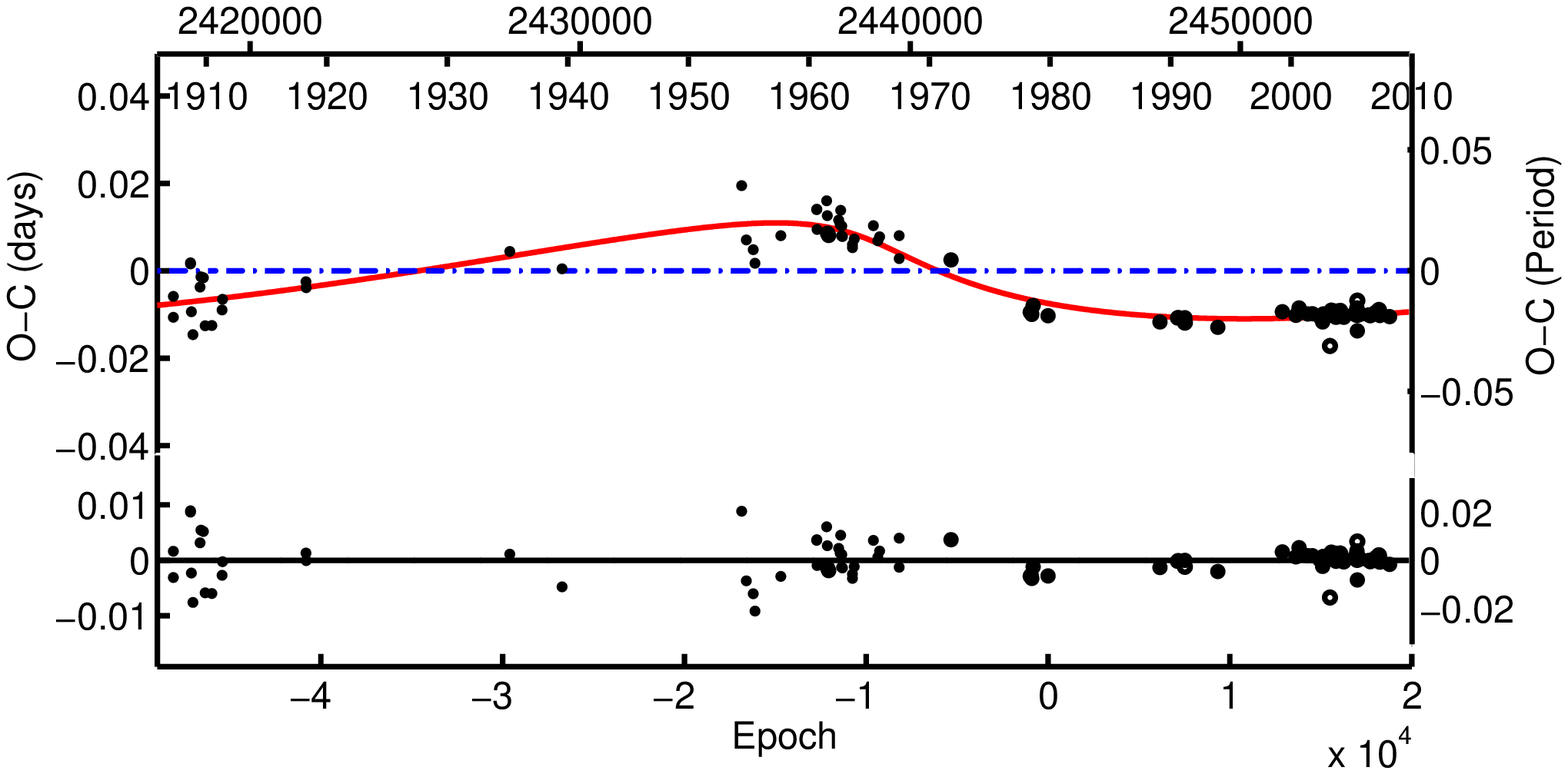}

\caption{The O-C diagram of EG Cep (left panel) and RZ Dra (right
panel) fitted by a LITE curve and a parabola (upper part) and the
total residuals after the subtraction of the whole solution (lower
part). The red solid line indicates the sum of the solutions,
while the blue dashed line corresponds to the parabola. The bigger
the symbol the bigger the weight assigned.}

\label{fig1}
\end{figure}

In order to analyze the O-C diagram of each system, the least
squares method with statistical weights has been used. The
construction of the O-C diagrams of EG Cep and RZ Dra are based on
a total of 258 and 91 times of minima taken from literature and
from our observations (see Table \ref{tab2}), respectively. For
both cases, the derived values of the O-C diagram solution are
listed in Table \ref{tab3}, and in Figure \ref{fig1} are shown the
fits on the data points and the O-C residuals after the
subtraction of the solution.

\begin{table}
\caption{The results of the O-C diagram analysis of EG Cep and RZ
Dra}
\label{tab3}

\begin{center}
\scalebox{0.7} {
\begin{tabular}{ccc}

\hline
\textbf{System}                                 & \textbf{EG Cep}  &   \textbf{RZ Dra}   \\
\hline
\textbf{Parameters of the EB}                   &                  &                     \\
\hline
$Min.~I$~[HJD]                                  & 2435956.5430~(3) &  2437181.4353~(2)   \\
P [days]                                        &   0.5446213~(1)  &   0.5508763~(1)     \\
$M_{1}$ + $M_{2}$~[$M_\odot$]                   &    1.0 + 0.84    &    1.62 + 0.66      \\
$c_{2}~(\times 10^{-10})$~[days/cycle]          &    0.1307~(1)    &     0.0349~(1)      \\
$\dot{P}~(\times 10^{-8})$~[days/year]          &    1.753~(1)     &     0.463~(1)       \\
$\dot{M}~(\times 10^{-9})$~[$M_{\odot}$/years]  &     16.72~(1)    &      3.70~(1)       \\
\hline
\textbf{Parameters of the $3^{rd}$ body}        &                  &                     \\
\hline
$P_3$~[yrs]                                     &     74.4~(9)     &     118.1~(1)       \\
$e_{3}$                                         &     0.59~(4)     &      0.72~(2)       \\
$M_{3,min}$~[$M_{\odot}$]                       &     0.15~(2)     &      0.226~(3)      \\
\hline
$\chi^{2}$                                      &     0.1538       &      0.0050         \\
\hline

\end{tabular}}
\end{center}
\end{table}

\section{Light Curve Analysis}
\begin{figure}[h]
\plottwo {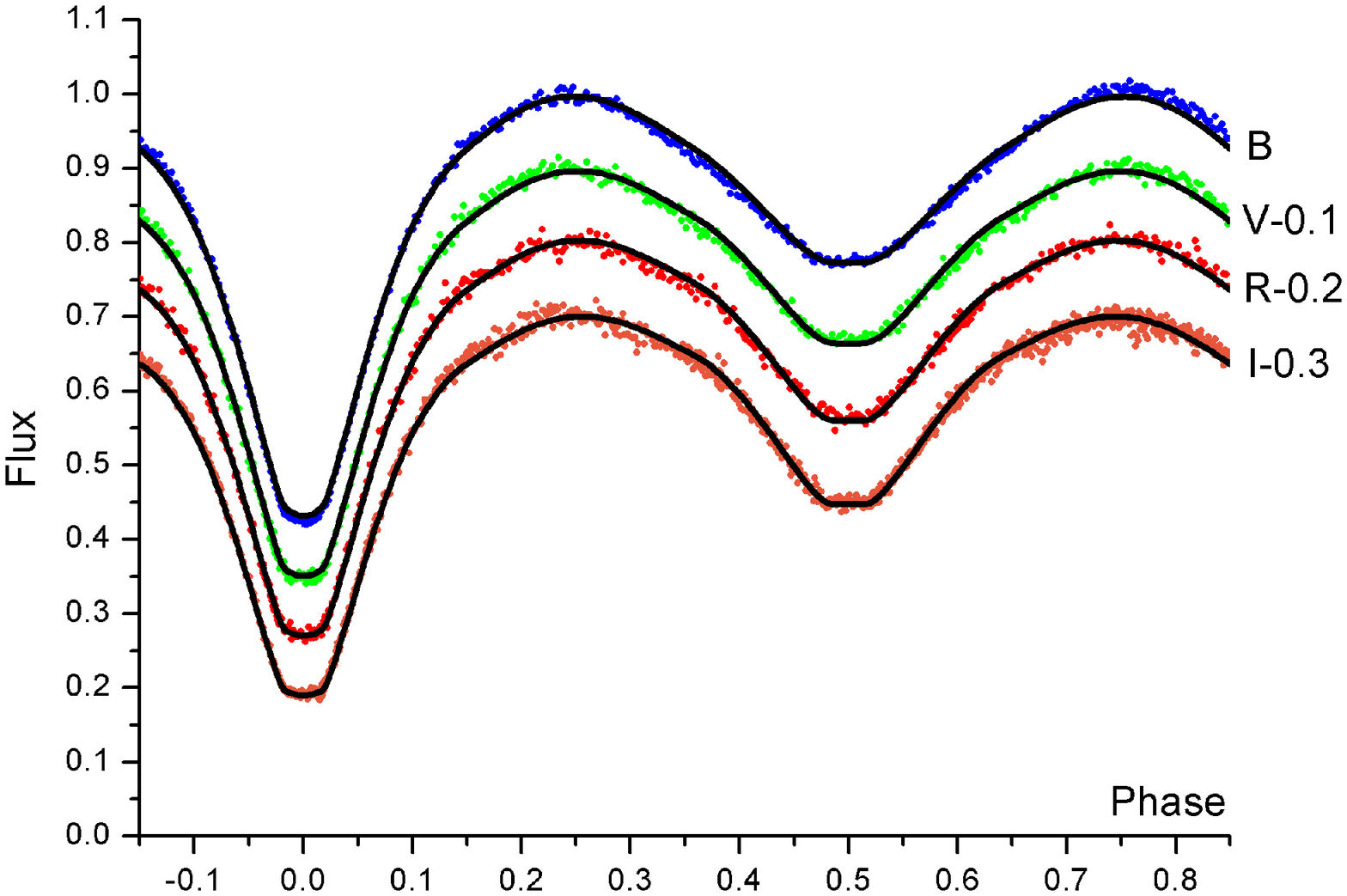}{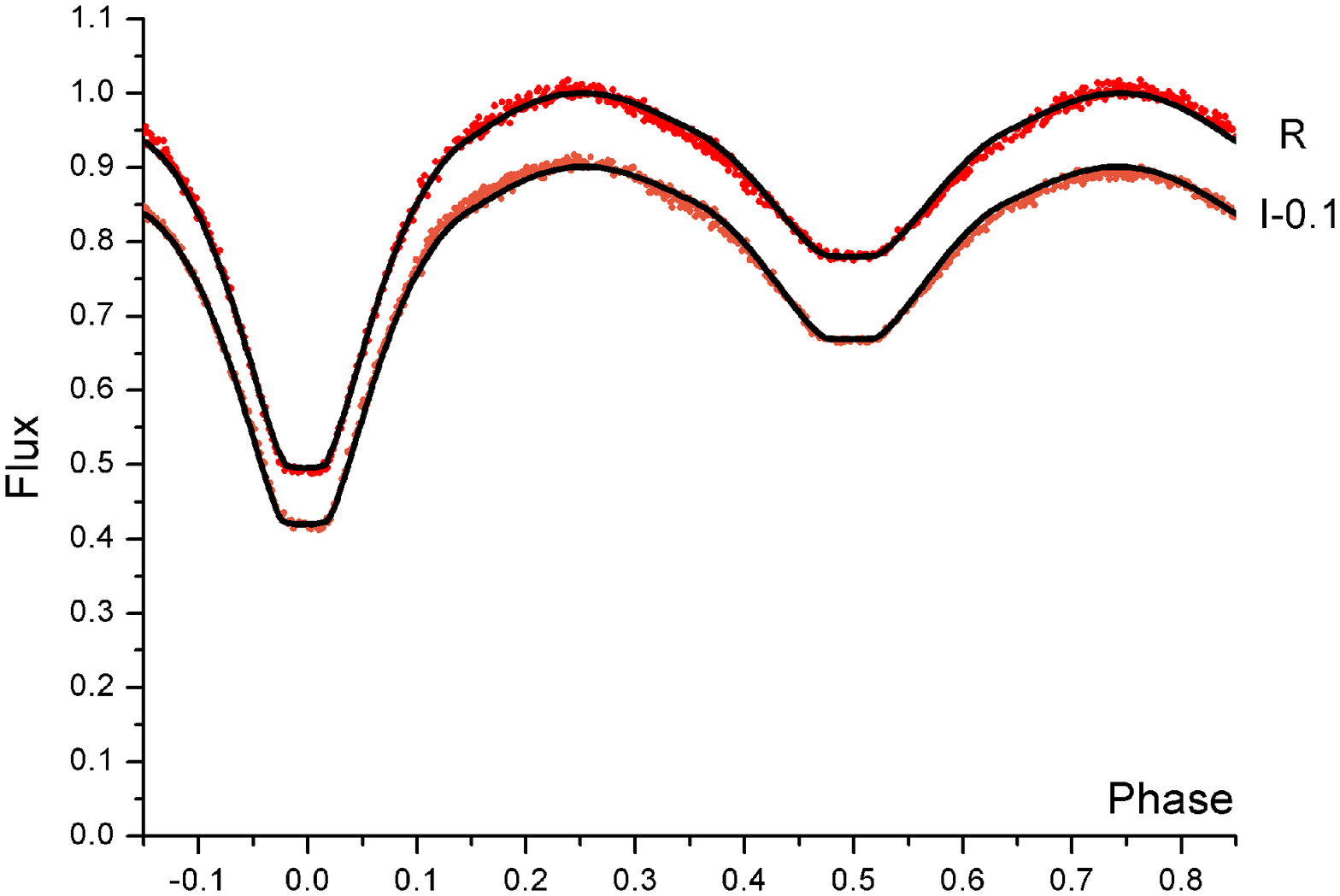}

\caption{Observed (colored points) and theoretical (solid lines)
LCs of EG Cep (left panel) and RZ Dra (right panel).}

\label{fig2}
\end{figure}

The light curves (hereafter LCs) have been analysed with the
Wilson Program (version 2003) \citep{WD71,W79,W90} and PHOEBE
software \citep{PZ05}. We applied the code in MODE 5, which solves
the LC of semi-detached eclipsing binaries, where the secondary
(cooler) component fills its Roche lobe, while the primary
(hotter) one is well inside its Roche lobe. For radial velocity
analysis of RZ Dra, we used 64 points data given by \citet{R00}.
The BVRI LCs of EG Cep, and the radial velocity and the RI LCs of
RZ Dra were used simultaneously for determination of the geometric
and physical elements of each system.\\
The LC solution is summarized in Tables \ref{tab4} and \ref{tab5}
for EG Cep and RZ Dra, respectively, and the theoretical and
observed LCs are illustrated in Figure \ref{fig2}.

\begin{table}
\caption{Physical and geometrical parameters of EG Cep}
\label{tab4}
\begin{center}
\scalebox{0.7} {
\begin{tabular}{cccc}

\hline
 \textbf{Parameter}   &\textbf{Value} &  \textbf{Parameter}   &  \textbf{Value}      \\
\hline
    $i$~[deg]         &    87.5~(2)   &   $q~(m_{2}/m_{1})$   &   0.449~(1)          \\
    $T_{1}^{*}$~[K]   &     8500      &       $T_{2}$~[K]     &   5792~(9)           \\
$L_{1V}/(L_{1}+L_{2})$&   0.650~(4)   &$L_{2V}/(L_{1}+L_{2})$ &   0.350~(4)          \\
    $r_{1(pole)}$     &  0.4092~(4)   &    $r_{2(pole)}$      &  0.2914~(2)          \\
    $r_{1(side)}$     &  0.4326~(5)   &    $r_{2(side)}$      &  0.3039~(2)          \\
    $r_{1(back)}$     &  0.4553~(6)   &    $r_{2(back)}$      &  0.3365~(2)          \\
   $\Omega_{1}$       &   2.859~(2)   & $(\sum ($res$)^{2})$  &   0.071              \\
\hline
$^{*}assumed$                                                                        \\
\end{tabular}}
\end{center}
\end{table}

\begin{table}
\begin{center}
\caption{Physical and geometrical parameters of RZ Dra}
\label{tab5} \scalebox{0.7}{
\begin{tabular}{cccc}

\hline
   \textbf{Parameter}   &\textbf{Value} &  \textbf{Parameter}   & \textbf{Value}\\
\hline
$R_{1}$~[$R_{\odot}$]   &       1.62    & $R_{2}$~[$R_{\odot}$] &    1.13       \\
$M_{1}$~[$M_{\odot}$]   &       1.617   & $M_{2}$~[$M_{\odot}$] &   0.656       \\
$M_{bol,1}$             &       2.84    &      $M_{bol,2}$      &    3.52       \\
$\alpha$~[${R_{\odot}}]$&     3.71~(2)  & $V_{\gamma}$~[km/sec] &   14.9~(3)    \\
$i$~[deg]               &      87~(1)   &   $q~(m_{2}/m_{1})$   &  0.406(2)     \\
$T_{1}^{*}$~[K]         &      8150     &       $T_{2}~[K]$     &  5531~(10)    \\
$r_{1(pole)}$           &   0.4157~(6)  &       $r_{2(pole)}$   &  0.2837~(3)   \\
$r_{1(side)}$           &   0.4399~(7)  &       $r_{2(side)}$   &  0.2957~(3)   \\
$r_{1(back)}$           &   0.4618~(3)  &       $r_{2(back)}$   &  0.3284~(3)   \\
$\Omega_{1}$            &   2.781~(3)   &  $(\sum ($res$)^{2})$ &    0.077      \\
\hline
$^{*}assumed$                                                                   \\
\end{tabular}}
\end{center}
\end{table}

\section{Search for Pulsations}    
In order to reveal any possible pulsation nature of RZ Dra and EG
Cep we subtracted the theoretical LCs from the observed ones in
order to remove the proximity effects (reflection and
eclipticity). The frequency analysis was made with PERIOD 04
software \citep{LB05} on the LC residuals, and we found no
evidence of pulsational behaviour either in EG Cep nor RZ Dra.

\section{Discussion and Conclusions}
The LC analysis of EG Cep and RZ Dra showed that both of them are
semi-detached systems with the secondary component filling its
Roche Lobe. The periodic variations of the orbital periods of
these systems could be explained by adopting the existence of a
tertiary component, while the steady increase of their period is
probably due to the mass transfer procedure. In contrast with the
O-C diagram solution, the LC analysis for both systems showed that
there is not a third light contribution in the total luminosity of
the system. This disagreement can be explained by taking into
account the small values of mass of the third body found in each
case. Finally, we could not detect any pulsation nature of the
primary components of both systems. So, more accurate data (by
using larger telescope and better CCD) in the future might reveal
possible pulsational behaviour.

\acknowledgements
This work has been financially supported  by the
Special Account for Research Grants No 70/4/5806 of the National \&
Kapodistrian University of Athens, Greece. The present work was made
using minima database: http://var.astro.cz/ocgate/. We thank P.
Zasche for providing the Matlab code to compute the results of the
O-C analysis.


\begin{thebibliography}{}

\bibitem[Kwee \& van Woerden(1956)]{KVW56}
Kwee, K., van Woerden, H. 1956, Bulletin of the astronomical
institutes of the Netherlands, 12, 464

\bibitem[Lenz \& Breger(2005)]{LB05}
Lenz, P. \& Breger, M. 2005, CoAst, 146, 53

\bibitem[Pr\v{s}a \& Zwitter(2005)]{PZ05}
Pr\v{s}a, A., \& Zwitter, T. 2005, ApJ, 628, 426

\bibitem[Rucinski et al.(2000)]{R00}
Rucinski, S.~M., Lu, W., \& Mochnacki, S.~W. 2000, AJ, 120, 1133

\bibitem [Soydugan et al.(2006)]{S06}
Soydugan, E., Soydugan, F., Demircan, O. \& \.{I}bano\v{g}lu, C.
2006, MNRAS, 370, 2013

\bibitem[Wilson \& Devinney(1971)]{WD71}
Wilson, R. E., \& Devinney, E. J. 1971, ApJ, 166, 605

\bibitem[Wilson(1979)]{W79}
Wilson, R. E. 1979, ApJ, 234, 1054

\bibitem[Wilson(1990)]{W90}
Wilson, R.~E. 1990, ApJ, 356, 613


\end{thebibliography}
\end{document}